\documentclass[prb,showpacs,amsmath,amssymb,nofootinbib]{revtex4}

\usepackage{bm}
\usepackage[dvipdfm]{graphicx}

\newcommand  {\eqn}[1]{(\ref{eqn:#1})}
\newcommand  {\Z}     {\mathbb{Z}}
\renewcommand{\(}     {\left(}
\renewcommand{\)}     {\right)}
\renewcommand{\_}[1]  {_{\rm #1}}

\begin{document}

\markboth{K. Kobayashi, T. Ohtsuki, and K. Slevin}
{Critical exponent for the quantum spin Hall transition in $\Z_2$ network model}

\title{Critical exponent for the quantum spin Hall transition in $\Z_2$ network model}

\author{K. Kobayashi}
 \email{k-koji@sophia.ac.jp}
\author{T. Ohtsuki}
\address{Department of Physics, Sophia University\\
Kioi-cho 7-1, Chiyoda-ku, Tokyo 102-8554, Japan}

\author{K. Slevin}
\address{Department of Physics, Graduate School of Science, Osaka University\\
Machikaneyama 1-1, Toyonaka, Osaka 560-0043, Japan}

\begin{abstract}
 We have estimated the critical exponent describing the divergence of the localization length at the metal-quantum spin Hall insulator transition.
 The critical exponent for the metal-ordinary insulator transition
in quantum spin Hall systems is known to be consistent with that of topologically trivial symplectic systems.
 However, the precise estimation of the critical exponent for the metal-quantum spin Hall insulator transition proved to be problematic because of the existence, in this case, of edge states in the localized phase.
 We have overcome this difficulty by analyzing the second smallest positive Lyapunov exponent instead of the smallest positive Lyapunov exponent.
 We find a value for the critical exponent $\nu=2.73\pm 0.02$ that is consistent with that for topologically trivial symplectic systems.

\end{abstract}


\maketitle

\section{Introduction}
 Onoda {\it et al.}\cite{OnodaAvishaiNagaosa} have questioned whether Anderson transitions in topologically non-trivial systems share the same critical properties as those in topologically trivial systems.
 Obuse {\it et al.},\cite{Obuse:Z2boundary,Obuse:nuSum} studied the metal-ordinary insulator transition in quantum spin Hall (QSH) systems and found that the value of the critical exponent is the same as that of topologically trivial symplectic systems.
 However, Obuse {\it et al.} only studied transitions to insulating phases without edge states leaving open the possibility 
that the critical exponent for the metal-QSH insulator transition might be different.
 Indeed, the critical conductance distributions have been found to be sensitive to the number of the edge states.\cite{Kobayashi:PGQSH}

 Here, we report an estimation of the critical exponent for the divergence of the localization length at the metal-QSH insulator transition.
 We find a value of the critical exponent that is consistent with that in the conventional symplectic class.
 Our result supports the conjecture that the metal-QSH insulator transition belongs to the conventional Wigner-Dyson symplectic class.

\subsection{Second smallest positive Lyapunov exponent}
 In order to estimate the critical exponents for the divergence of the localization length, 
we have performed a finite size scaling analysis of numerical data for Lyapunov exponents.\cite{MacKinnonKramer1,MacKinnonKramer2,KramerMacKinnon,Kramer:bookFSS}
 This involves the estimation of the Lyapunov exponents for quasi-one-dimensional systems with effectively infinite length and given cross sections.
 For extrapolation to the two dimensional limit that is of interest here, a strip with a cross section $L$ is considered.
 For ordinary metal-insulator transitions, 
the standard approach is to analyze the scaling of the dimensionless quantity
   \begin{equation}\label{eqn:Gamma1}
    \Gamma = \Gamma_1 = \gamma_1 L,
   \end{equation}
which is equal to the product of the smallest positive Lyapunov exponent $\gamma_1$ (precisely defined below)
and the cross section. 
(This quantity is the inverse of the so-called MacKinnon-Kramer parameter.)
 In a metallic phase $\Gamma$ usually increase with $L$ and in a localized phase usually decreases with $L$.
 At the transition point between these phases $\Gamma$ becomes independent of $L$ 
and a common crossing point of curves with different $L$ is visible on the appropriate graph.
 However, when applying this method to the  metal-QSH insulator transition we run into a problem:
$\Gamma$ increases in both the metallic and QSH insulating phases and no common crossing point is seen (Fig.~\ref{fig:LE1st}).
 This problem, which makes a very precise scaling analysis difficult, occurs because of the existence of edge states in the QSH insulating phase.
 We have overcome this problem by analyzing the scaling of the quantity
   \begin{equation}\label{eqn:Gamma2}
    \Gamma_2 = \gamma_2 L,
   \end{equation}
which involves the second smallest positive Lyapunov exponent.
 This Lyapunov exponent is much less affected by these edge states and a common crossing point is recovered (Fig.~\ref{fig:LE2nd}). 
 It has been demonstrated\cite{Slevin:01} for the Anderson transition in three dimensional systems in the orthogonal symmetry class that the critical exponent obtained from the scaling of higher Lyapunov exponents is the same as that obtained from the scaling of the smallest positive exponent.

\section{Calculation of Lyapunov exponents}

\subsection{Model} \label{sec:model}
\begin{figure}[tb]
  \begin{center}
  \begin{tabular}{cc}
   \includegraphics[width=60mm]{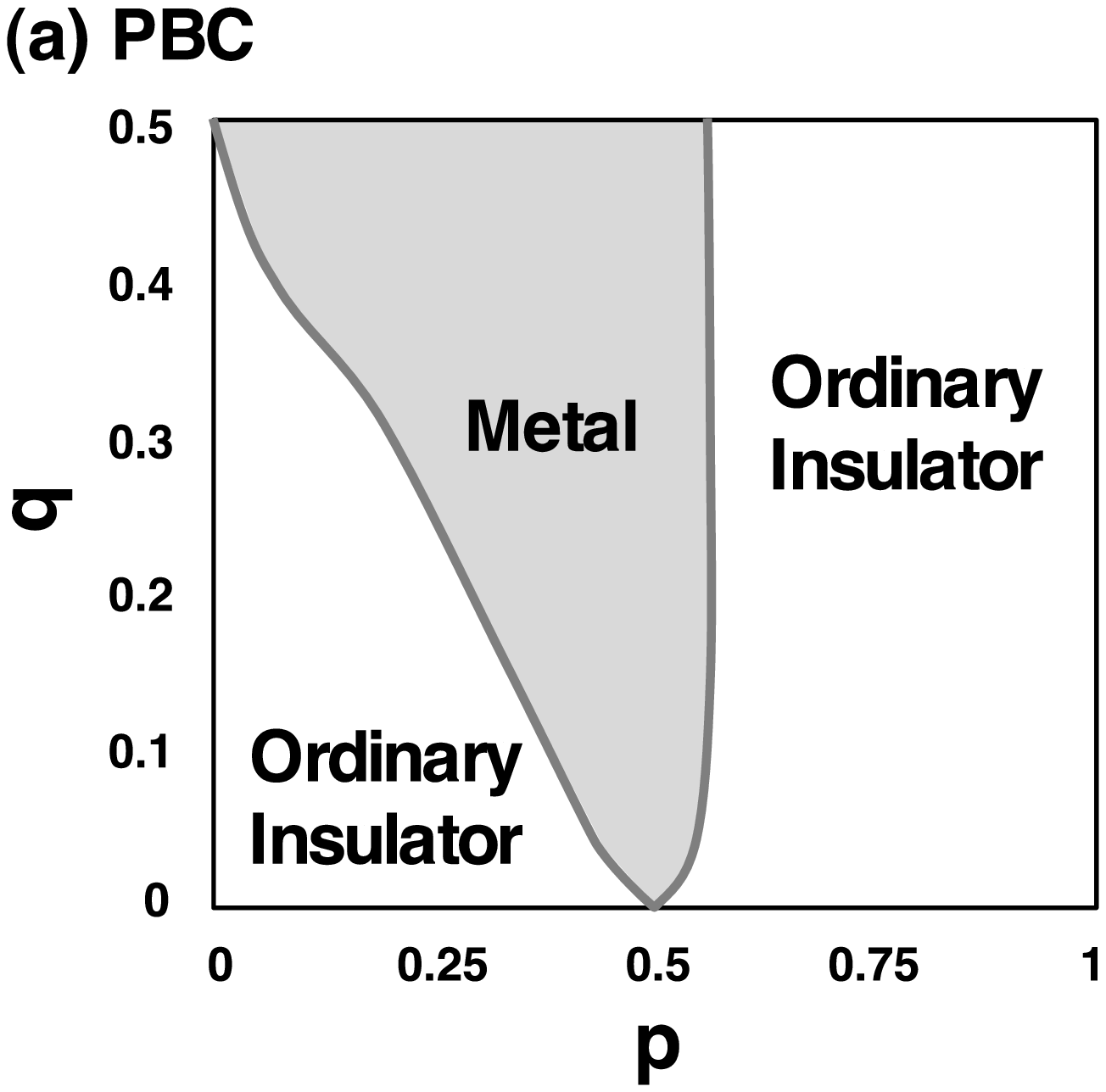} &
   \includegraphics[width=60mm]{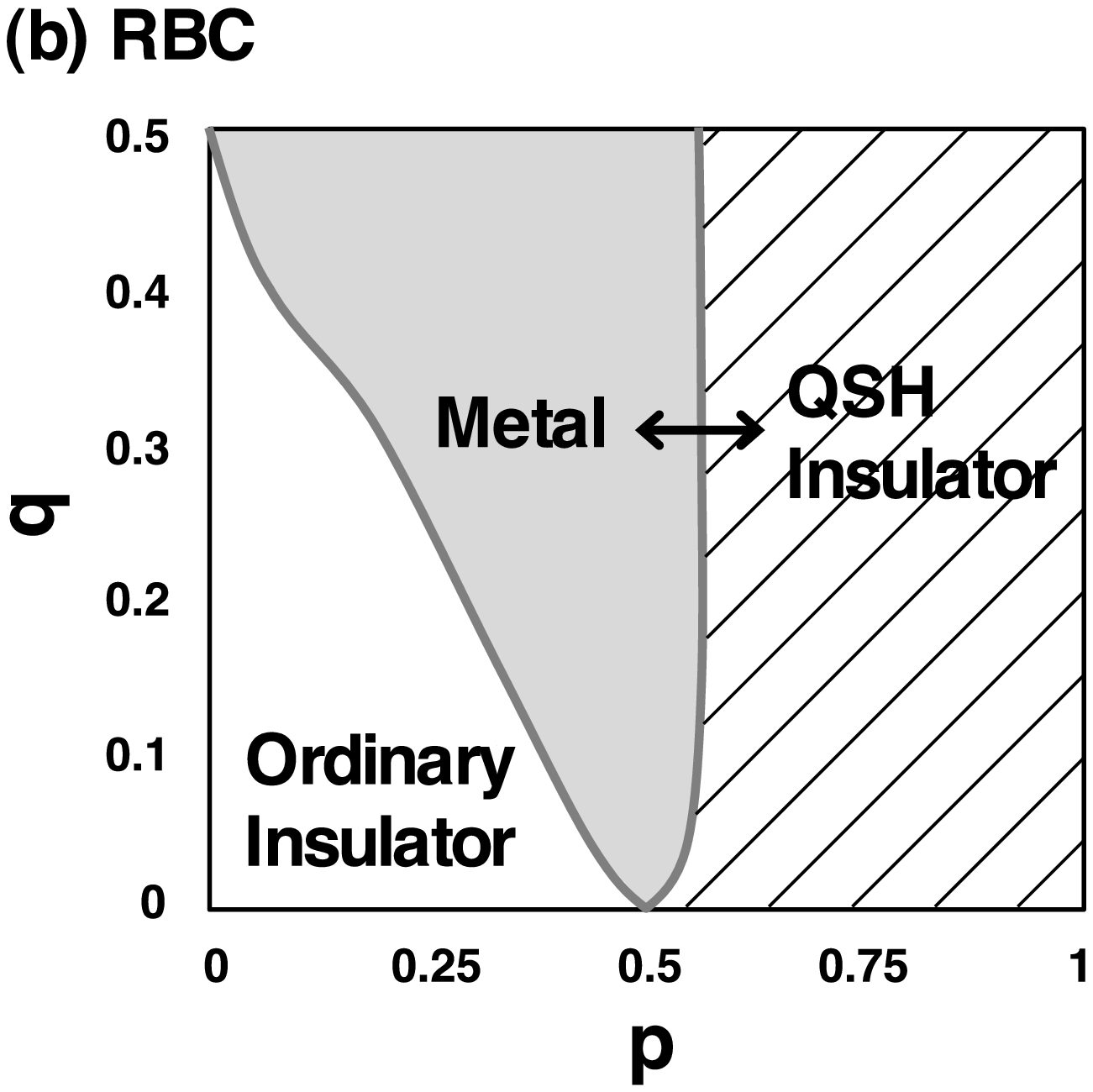}
  \end{tabular}
  \end{center}
  \vspace{-3mm}
  \caption{
     Phase diagram of $\mathbb{Z}_2$ network model with (a) PBC and (b) RBC.
     When RBC are imposed, 
     the insulating phase without edge states appears for small $p$,
     the quantum spin Hall insulating phase appears for large $p$, and 
     a metallic phase appears between the two insulating phases.
     The arrow indicates the range of the parameter considered in this paper.
  }
  \label{fig:phaseDiag}
 \end{figure}

 To describe the QSH system, we use a $\mathbb{Z}_2$ quantum network model.\cite{Obuse:QSHnwm,Obuse:QSHnwm2,Ryu:Z2nwm}
 The $\Z_2$ network model has two controlling parameters, $p$, the tunneling probability (related to the chemical potential), and $q$, the spin-mixing probability (related to the spin-orbit interaction strength).
 For finite $q$, the $\mathbb{Z}_2$ network model 
exhibits a metallic phase sandwiched between two insulating phases.
 When periodic boundary conditions (PBC) are imposed,
both insulating phases correspond to the ordinary insulating phase [Fig.~\ref{fig:phaseDiag}(a)].
 On the other hand, when reflecting boundary conditions (RBC) are imposed,
the insulating phase located for larger $p$ exhibits edge states
and becomes the QSH insulating phase [Fig.~\ref{fig:phaseDiag}(b)].
 Thus, the $\mathbb{Z}_{2}$ network model with RBC shows two types of transition:
the metal-ordinary insulator transition and the metal-QSH insulator transition.\footnote{
 The direct transition from the ordinary to QSH insulator occurs only for $q\to0$.
 In this limit, the transition point belongs to the same universality class as the integer quantum Hall system.}

 The cross section of the network is measured in terms of the number of links;
a slice of width $L$ contains $L$ links, and each link has a spin degree of freedom.
 In this paper we focus on the case of $L$ even with RBC.
 (For $L$ odd, perfectly conducting channels appear.\cite{AndoSuzuura:PCC,Kobayashi:LT25})

\subsection{Lyapunov exponents}
 We consider a quasi-one-dimensional system.
 The Lyapunov exponents are estimated using a transfer matrix method.
 The transfer matrix $T_x$ relates the current amplitudes $\psi_{x}$ on the slice $x$ to those on the next slice, 
   \begin{align}
      \psi_{x+1} = \bm{T}_{x}\psi_{x}.
   \end{align}
 Consider the matrix ${\cal T}^\dag{\cal T}$, where ${\cal T}$ is the transfer matrix product,
   \begin{align}
    {\cal T} = \bm{T}_{M} \cdots \bm{T}_{2} \bm{T}_{1}.
   \end{align}
 The matrix ${\cal T}^\dag{\cal T}$ is Hermitian and positive definite (because ${\cal T}$ is invertible) 
and so its eigenvalues are real and positive.
 They also occur in reciprocal pairs because of current conservation.
 From each eigenvalue a Lyapunov exponent is defined by the limit
   \begin{align} \label{eqn:LEith}
    \gamma = \lim_{M\to\infty} \frac{\ln\lambda}{2M},
   \end{align}
where $\lambda$ is an eigenvalue of ${\cal T}^\dag{\cal T}$.
 In QSH systems, states are doubly degenerate (Kramers degeneracy) and the same degeneracy occurs in the Lyapunov exponents.
 Keeping this in mind, and putting the exponents in decreasing order, we number them as
   \begin{align}
    \gamma_{L/2} > \gamma_{L/2-1} > \cdots > \gamma_{2} > \gamma_{1} > 0 > -\gamma_{1} > -\gamma_{2} > \cdots > -\gamma_{L/2}\, .
   \end{align}
 Here, it is to be understood that each exponent is doubly degenerate.

 The calculation of the Lyapunov exponent Eq.~\eqn{LEith} is terminated at finite $M$ 
when the target Lyapunov exponent has converged to within a specified precision.
 Note that higher Lyapunov exponents converge more quickly than lower Lyapunov exponents.

\section{Results}

\subsection{Finite size scaling}
 We consider a $\Z_2$ network model with strip geometry with RBC in transverse direction 
and the spin-mixing parameter set to $q=0.309$.
 For each $\Gamma$, we assume the scaling formula,
   \begin{align} \label{eqn:scaling}
    \hspace{-5mm}
      \Gamma(L,p) = F((p-p\_c) L^{1/\nu},  (p-p\_c) L^{y\_{irr}}),
   \end{align}
where $y\_{irr}(<0)$ is the exponent for the irrelevant correction to scaling.
 We expand the scaling function around the QSH transition point $p=p\_c$ as
   \begin{align} \label{eqn:scaling_ex}
    \hspace{-5mm}
      \Gamma(L,p) &= \Gamma\_c
                    + \sum^{n_c}_{j=1} c_j\! \(u L^{1/\nu}\)^j
                    + \sum^{m}_{i=1} \sum^{n_d(i)}_{j=0} d_j(i)\! \(u L^{1/\nu}\)^j \(v L^{y\_{irr}}\)^i,
   \end{align}
where $u$ is the relevant scaling variable,
   \begin{align} \label{eqn:u}
      u &= \sum^{n_a}_{j=1} a_j \(\frac{p-p\_c}{p\_c}\)^j, \quad a_1 = 1,
   \end{align}
and $v$ the irrelevant scaling variable,
   \begin{align} \label{eqn:v}
      v &= \sum^{n_b}_{j=0} b_j \(\frac{p-p\_c}{p\_c}\)^j, \quad b_0 = 1.
   \end{align}

\begin{figure}[tb]
  \begin{center}
   \begin{tabular}{c}
    \includegraphics[width=100mm]{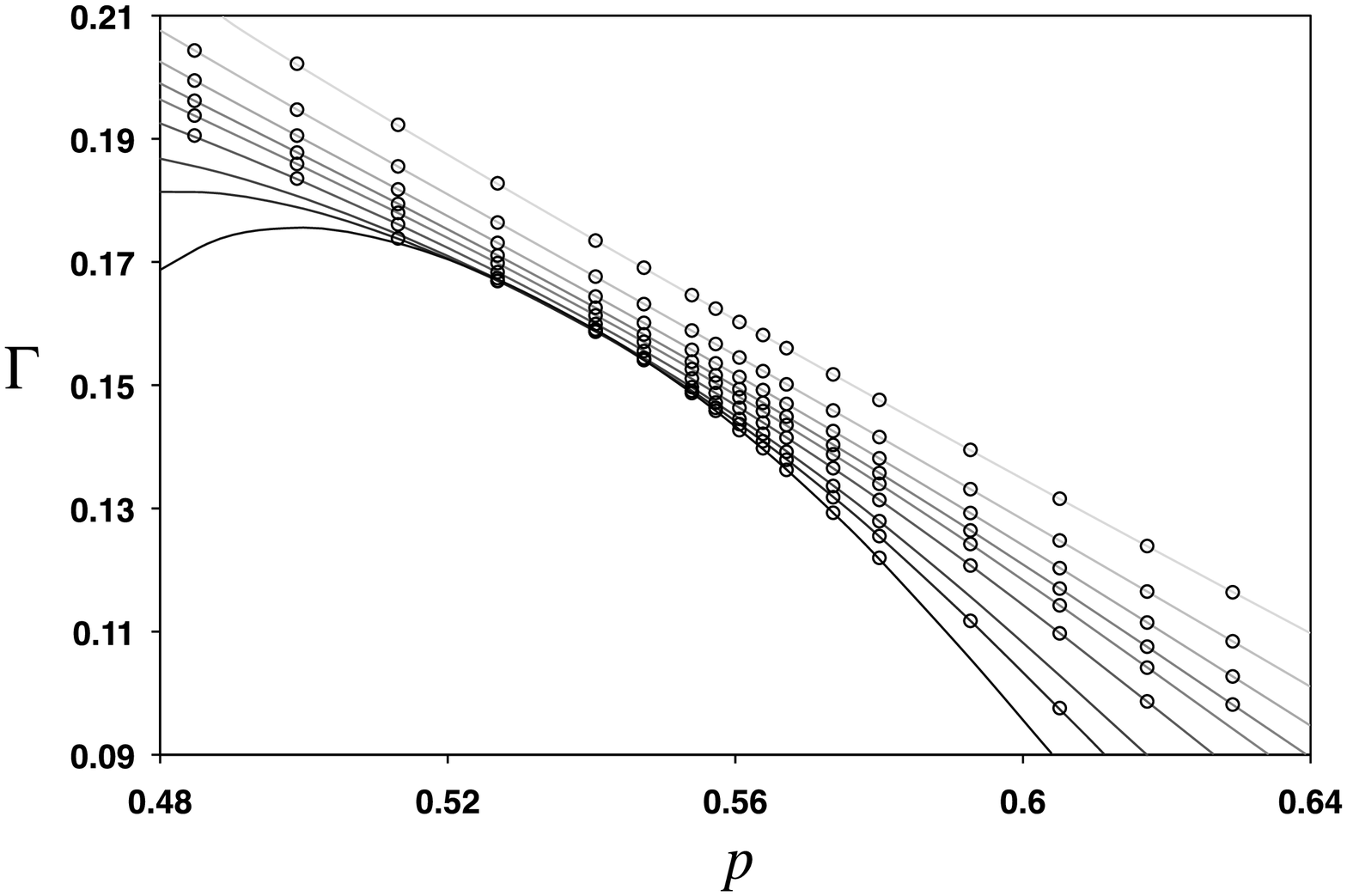}
   \end{tabular}
  \end{center}\vspace{-4mm}
  \caption{
     $\Gamma$ of Eq.~\eqn{Gamma1} versus tunneling probability 
     for the $\Z_2$ network model near the metal-QSH insulator transition.
     The lines are a finite size scaling fit, with different lines corresponding to different cross sections ($L=16,24,32,40,48,64,96,128,192$).
  }
  \label{fig:LE1st}
 \end{figure}

 First, we analyze $\Gamma$, which is calculated from the smallest positive Lyapunov exponent (see Fig.~\ref{fig:LE1st}).
 A fit of the numerical data to Eq.~\eqn{scaling_ex} with $n_a\!=\!3,\, n_b\!=\!1,\, n_c\!=\!5,\, n_d(1)\!=\!5,\, n_d(2)\!=\!2$ and taking $p\_c,\Gamma\_c,a_j,b_j,c_j,d_j(i),\nu,$ and $y\_{irr}$ as fitting parameters, yielded
   \begin{align} \label{eqn:fitted_1st}
    \hspace{-5mm}
      p\_c     \!=\! 0.561 \!\pm\! 0.001, \ 
      \Gamma\_c \!=\! 0.1400 \!\pm\! 0.0004, \ 
      \nu      \!=\! 2.64  \!\pm\! 0.06, \ 
      y\_{irr} \!=\! -0.94 \!\pm\! 0.04 \,.
   \end{align}
 The precision of the data for the smallest positive Lyapunov exponents is better than $0.03\%$ for $L=16,24,32,40,48,64$ and $0.05\%$ for $L=96,128,192$, 
and goodness of fit is $0.2$.
 Nevertheless, the precision of the estimate of the critical exponent $\nu$ is not sufficient to distinguish it from the critical exponent of the quantum Hall transition, $\nu=2.59\pm 0.01$.\cite{Slevin:IQHnu}

\begin{figure}[htb]
  \begin{center}
   \begin{tabular}{c}
    \includegraphics[width=100mm, bb=0 0 661 440]{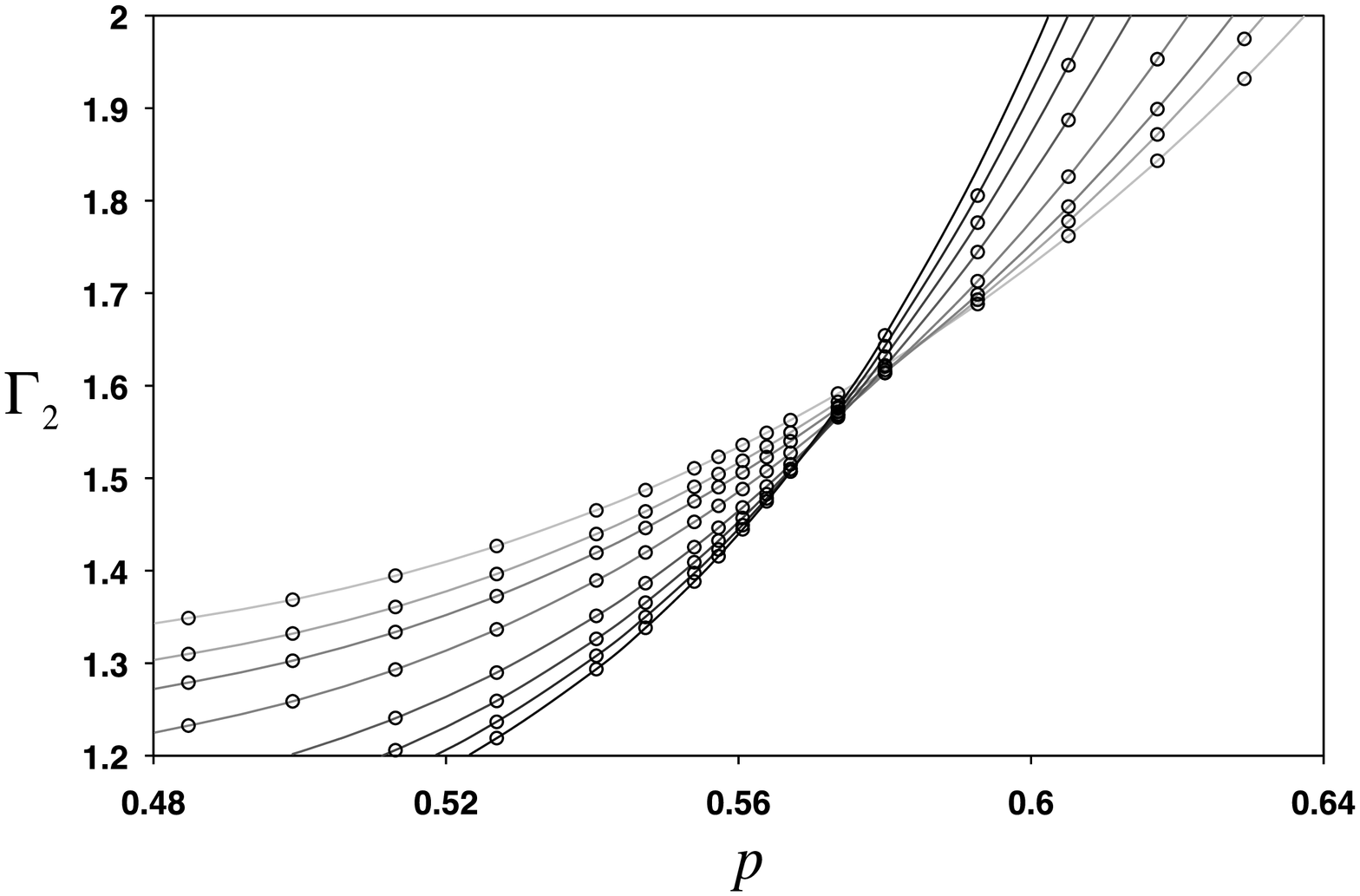}
   \end{tabular}
  \end{center}\vspace{-4mm}
  \caption{
     $\Gamma_2$ of Eq.~\eqn{Gamma2} versus tunneling probability 
     for the $\Z_2$ network model near the metal-QSH insulator transition.
     The lines are a finite size scaling fit, with different lines corresponding to different cross sections ($L=32,40,48,64,96,128,160,192$).
  }
  \label{fig:LE2nd}
 \end{figure}

 Next we analyze $\Gamma_2$, which is calculated from the 2nd smallest positive Lyapunov exponent (see Fig.~\ref{fig:LE2nd}).
 We see a clear common crossing point of the curves separating the metallic and topological insulating phases.
 A fit of the numerical data to Eq.~\eqn{scaling} with $n_a\!=\!2,\, n_b\!=\!1,\, n_c\!=\!4,\, n_d(1)\!=\!4,\, n_d(2)\!=\!2$ yielded
   \begin{align} \label{eqn:fitted_2nd}
    \hspace{-5mm}
      p\_c     \!=\! 0.562 \!\pm\! 0.001, \ 
      \Gamma\_c \!=\! 1.429 \!\pm\! 0.004, \ 
      \nu      \!=\! 2.73 \!\pm\! 0.02, \ 
      y\_{irr} \!=\! -0.95 \!\pm\! 0.02 \,.
   \end{align}
 The precision of the data for the 2nd Lyapunov exponents is better than $0.01\%$, 
and goodness of fit is $0.7$.
 This result is in good agreement with that obtained 
by Asada {\it et al.}\cite{Asada:nu} for SU(2) model with PBC, $\nu = 2.746 \pm 0.009$, 
and by Obuse {\it et al.}\cite{Obuse:Z2boundary} for $\Z_2$ network model with RBC at the metal-ordinary insulator transition, 
$\nu = 2.88 \pm 0.04$.
 It is also clearly different from the exponent for the quantum Hall transition.


\section{Summary}
 We have reported an estimate the critical exponent for the divergence of the localization length 
at the metal-quantum spin Hall insulator transition.
 By analyzing the scaling of the 2nd smallest positive Lyapunov exponent, 
we have estimated the critical exponent to be $\nu=2.73\pm0.02$.
 Our result shows that this critical exponent is insensitive to the topological property of the insulating phase.
 This is in sharp contrast to the critical conductance distribution,\cite{Kobayashi:PGQSH} 
which is sensitive to the type of transition, {\it i.e.}, the presence or absence of edge states in the adjacent insulating phase.

 Analysis of the scaling of higher Lyapunov exponents may also be useful in the study of other Anderson transitions 
where conducting edge or surface states occur in the insulating phase, such as, for example, in topological insulators.


\section*{Acknowledgments}
 This work was supported by KAKENHI No. 23$\cdot$3743 and KAKENHI No. 23540376.

\end{document}